\documentclass[12pt]{article}
\begin{document}
\title{
Bose-Fermi Equivalence in Three Dimensional Non-commutative
Space-Time}
\author{Ajith K M\thanks{ph01ph13@uohyd.ernet.in}, E. Harikumar\thanks{harisp@uohyd.ernet.in} and M. Sivakumar\thanks{mssp@uohyd.ernet.in}\\
School of Physics, University of Hyderabad,\\
Hyderabad, 500046, India.\\
} \maketitle
\begin{abstract}
We study the Fermionisation of Seiberg-Witten mapped action (to order $\theta$) of the $\lambda\phi^{4}$ theory coupled
minimally with U(1) gauge field governed by Chern-Simon action. Starting from the corresponding partition function we derive non-perturbatively (in coupling constant) the partition
function of the spin $\frac{1}{2}$ theory following
Polyakov spin factor formalism. We find the dual interacting
fermionic theory is non local. This feature
persist also in the limit of vanishing self coupling. In
$\theta\to 0$ limit, the commutative result is regained.
\end{abstract}
\newpage
\section{Introduction}
Non commutative (NC) space-time\cite{rev} has gained a lot of
interest in recent time due to (i) relevance to quantum aspects of
gravity (ii) as a regularization in field theory (iii) certain limit
of the string theory. Well studied among them is the Moyal
space-time whose co-ordinates obey
\begin{equation}
[x_{\mu},x_{\nu}]_{\ast}=i\theta_{\mu\nu}
\end{equation}
In Moyal space-time, the usual product is replaced by the  $\ast$
product defined as
 \begin{equation}
 f(x)\ast
 g(x)=e^{\frac{i}{2}\theta^{ij}\partial^{x}_{i}\partial^{y}_{j}}f(x)g(y)|_{x=y}
\end{equation}
and $\theta_{\mu\nu}$ is a constant parameter characterizing the
noncommutativity. Field theories on such space-time have several
interesting features which are distinct from the models on the
commutative space-time. These includes UV/IR mixing,  novel
topological soliton solutions, twisted symmetries etc. Gauge
theories in NC space time can be mapped using Seiberg-Witten (SW)
map \cite{sw} to gauge theories in commutative space time. In this
note we study Bose-Fermi equivalence of scalar field theory in 2+1 dimensional NC space-time. Bose-Fermi transmutation
is studied for the NC theories  after re expressed in the
commutative space-time using SW map(keeping up to order $\theta$
terms.) Bosonisation of Fermionic theories (and vice versa)have
been well studied in commuative space-time. Polyakov proposed a
study of Bose-Fermi equivalence in 2+1 dimension using
Chern-Simon gauge action$\cite{poly}$. It was shown that the
expectation value of the Wilson loop averaged over Chern-Simon(CS) gauge action
( and for a suitable coefficient) is given by
\begin{eqnarray}
<e^{i\oint_{c} Adx}>_{CS}=e^{i\pi W(C)} \label{poly}
\label{writh}
\end{eqnarray}
where $W(C)$ is the writhe of the space curve. It was also shown
\cite{poly,jacob} that $W(C)$ =$\Omega(C)+(2k+1), k\in Z$. Here
$\Omega(C)$, known as Polyakov spin factor, represent the solid
angle subtended by the tangent to the curve C on a unit sphere.
This is also related to the overlap of spin coherent states and
forms the symplectic 2-form of SU(2) group manifold which is
$S^{2}$. The odd integer (2k+1) has been shown to be related to
statistics. The expression in the Eqn.(\ref{writh}) has been applied, to derive
fermionic theory from scalar field coupled with $U(1)$ gauge field governed by Chern-Simon action by one of us \cite{sssmpl,pssmpl}. Interesting feature of this approach is that, it is non-perturbative in coupling constant.
In this work we apply this approach to NC Space-time. 

Bose-Fermi equivalence can be seen as duality equivalence. Duality
aspects of NC field theories have been well studied
$\cite{trg,sg,botta,cw,davi}$ in recent times. There were
many studies using different approaches to generalise the known
duality equivalence between Maxwell-Chern-Simon theory(MCS) to
Self-dual model(SD) in NC spaces. This was studied in
$\cite{sg,botta}$ using master action method but with different
conclusions. By applying a dual projection procedure to NCSD model, a
dual model was constructed in $\cite{cw,davi}$ and was shown to be
different from NCMCS theory. In $\cite{voreh}$, using a different
approach dual of SW mapped NCMCS was obtained and shown to be
different from SW mapped NC (St\"uckelberg compensated) SD model.
These investigations showed that the duality relation present in
the commutative space time need not carry forward to NC
space-time. Bosonization in two and three dimensional NC
space-time has been studied in \cite{sg1,sch,das}.
 Hence, it is interesting to
investigate bosonisation in NC space time following the Polyakov
approach.

 In this work we study Fermionisation of 
$\lambda\phi^{4}$ theory coupled to $U_{\ast}(1)$ gauge field governed by Chern-Simon action 
in NC space-time. We apply SW map to re-express the theory interms of fields in commutative space-time 
keeping terms up to the first order in $\theta$, and apply the methods devolepd in \cite{sssmpl,pssmpl}.  We derive
the Fermionic partition function, exact in self coupling. The dual
Fermionic theory obtained is nonlocal, interacting
theory. We see that the Fermionic mass term does not get $\theta$ correction.

\section{Scalar field in fundamental representation of  $U_{\ast}(1)$}
In this section, we consider Fermionisation of self interacting
scalar field theory in the fundamental representation of
$U_{\ast}(1)$. We start with the massive complex scalar field in
fundamental representation, coupled to a Chern-Simon term in noncommutative
 Euclidean space described by
\begin{eqnarray}
\hat S_{\phi}&=&\int{d^{3}x }\big[(\hat D^{\mu}\hat\phi)\ast(\hat
D_{\mu}\hat\phi)^{\dagger}+m^{2}\hat\phi\ast\hat\phi^{\dagger}-\lambda(\hat\phi^{\dagger}\ast\hat\phi)\ast(\hat\phi^{\dagger}\ast\hat\phi)\nonumber\\&-&\frac{i}{4\pi}\epsilon_{\mu\nu\lambda}(\hat
A_{\mu}\partial_{\nu}\hat A_{\lambda}+\frac{2i}{3}\hat A_{\mu}\hat
A_{\nu}\hat A_{\lambda})\big] \label{fact}
\end{eqnarray}
 In the above  action hated fields are functions of non
commutative (NC) co-ordinates. The covariant derivative is defined
by
$$\hat D_{\mu}\hat\phi=\partial_{\mu}\hat\phi-i\hat
A_{\mu}\ast\hat\phi.$$ Using SW map we rewrite the action in Eqn.(\ref{fact}) in terms of commutative fields  and $\theta$. For this we use the SW solution for $\hat \phi$ and $\hat A_{\mu}$  
\begin{eqnarray}
\hat A_{\mu}&=&A_{\mu}-\frac{1}{2}\theta^{\alpha\beta}A_{\alpha}(\partial_{\beta}A_{\mu}+F_{\beta\mu})\\
\label{sw}
 \hat
 \phi&=&\phi-\frac{1}{2}\theta^{\alpha\beta}A_{\alpha}\partial_{\beta}\phi.
\end{eqnarray}
to order $\theta$ \cite{VOR}. Using this, from Eqn.(\ref{fact}) we get (to order $\theta$)
\begin{eqnarray}
S_{\phi}&=&\int{d^{3}}x[ D^{\mu}\phi
(D_{\mu}\phi)^{\dagger}-y^{\mu\nu}D_{\mu}\phi(D_{\nu}\phi)^{\dagger}\nonumber\\&+&m^{2}(1+y^{\mu}_{~\mu})\phi\phi^{\dagger}-\lambda(\phi\phi^{\dagger})^{2}
(1-\frac{1}{2}\theta^{\alpha\beta}F_{\alpha\beta})-\frac{i}{4\pi}\epsilon_{\mu\nu\lambda}
A_{\mu}\partial_{\nu} A_{\lambda}]
\end{eqnarray}
where
$$y^{\mu\nu}=\frac{1}{2}(\theta^{\mu\alpha}F_{\alpha}^{~\nu}+\theta^{\nu\alpha}F_{\alpha}^{~\mu}+\frac{1}{2}\eta^{\mu\nu}\theta^{\alpha\beta}F_{\alpha\beta}).$$
 Note that we have used the fact that NC Chern-Simon term goes to commutative Chern-Simon term under SW map \cite{GS}. The coefficient of Chern-Simon terms is chosen sothat dual theory is that of spin-$\frac{1}{2}$ fermion. Before integrating the scalar fields we linearise $\lambda$ term in the above and re-express this term using
 Hubbard-Stratnovich field $\chi$, 
 \begin{equation}
\lambda(\phi\phi^{\dagger})^{2}
(1-\frac{1}{2}\theta^{\alpha\beta}F_{\alpha\beta})=-\chi(x)^{2}+2\sqrt{\lambda}\chi(x)(\phi\phi^{\dagger})(1-\frac{1}{4}\theta^{\alpha\beta}F_{\alpha\beta})
 \end{equation}
 \section{Scalar field integration}
  The Euclidean path integral with the above
action is given by
\begin{eqnarray}
Z=\int{D\phi D\phi^{\dagger}DA DB DC D\chi
~e^{-S_{0}}~e^{-(\int{d^{3}x[\frac{i}{4\pi}\epsilon_{\mu\nu\lambda}
A_{\mu}\partial_{\nu}
A_{\lambda}]+iC_{\mu\nu}(y^{\mu\nu}+B^{\mu\nu})+\chi(x)^{2}})}}
\end{eqnarray}
 $$S_{0}=\int{d^{3}x}\Big[ D^{\mu}\phi
(D_{\mu}\phi)^{\dagger}+B^{\mu\nu}D_{\mu}\phi(D_{\nu}\phi)^{\dagger}+[\tilde
m^{2}(1-B^{\mu}_{~\mu})]\phi\phi^{\dagger}\Big]$$
 $C_{\mu\nu}$ and $B_{\mu\nu}$ were introduced
 to linearize the  $\theta$ depended coupling of A field to scalar field and and we use $\tilde
 m^{2}(x)=m^{2}-2\sqrt{\lambda}~\chi(x)$.

 After integrating the $\phi$ and $\phi^{\dagger}$  fields we get
partition function as
\begin{eqnarray}
 Z=\int{DA DB DC~
 e^{-ln \,det{\cal \,R}}~ e^{-i(\int{C_{\mu\nu}(y^{\mu\nu}_{\theta}+B^{\mu\nu})+\chi^{2})+\frac{i\lambda}{4\pi^{2}}\epsilon_{\mu\nu\lambda}
A_{\mu}\partial_{\nu} A_{\lambda}})}}
\end{eqnarray}
where the operator ${\cal R}$~is given by
 \begin{eqnarray}
{\cal
R}&=&(-(\delta^{\mu\nu}+B^{\mu\nu})D_{\mu}D_{\nu}-(D_{\mu}B^{\mu\nu})D_{\nu}+V)\nonumber\\
 \hbox{and}~~~~V(x)&=&[\tilde
m^{2}(1-B^{\mu}_{~\mu})].
 \label{v}
\end{eqnarray}
  We can use the heat
kernel representation of the logarithm of determinant\cite{heat}, treating
$\cal R$ as the Hamiltonian, i.e,
\begin{eqnarray}
ln~ det~~{\cal
R}=\int_{\frac{1}{\Lambda^{2}}}^{\infty}{\frac{d\alpha}{\alpha}~Tr
~e^{-\alpha{\cal R}}}
\end{eqnarray}
 Applying the standard path integral
method to this gauge invariant ``Hamiltonian", ${\cal R}$ we obtain
\begin{eqnarray}
 ln Det {\cal R}=\int_{\frac{1}{\Lambda^{2}}}^{\infty}{\frac{d\alpha}{\alpha}\int_{x(\alpha)=x(0)}Dx(\tau){e^{-\int_{0}^{\alpha}{d\tau
[{\cal H}+\frac{1}{2}\dot
x^{\mu}\partial^{\rho}B_{\rho\mu}-i\dot x^{\mu}A_{\mu}]}}}}
\end{eqnarray}
In the above the measure $Dx(\tau)=
(4\pi\epsilon)^{\frac{-3N}{2}}\prod_{i=0}^{N-1}d^{3}x$, and
${\cal H} =\frac{1}{4}(M^{\mu\nu})^{-1}\dot x_{\mu}\dot
x_{\nu}+V(x(\tau))$ with
$M^{\mu\nu}=(\delta^{\mu\nu}+B^{\mu\nu})$. Here we take
$\epsilon\rightarrow\frac{1}{\Lambda^{2}}$. Also we omitted terms
quadratic in $B_{\mu\nu}$ as they are of order $\theta^{2}$  which can be seen
by integrating C-field. After expanding the $e^{-ln~ det ~{\cal
R}}$ in power series, the partition function becomes
\begin{eqnarray}
&&Z=\int{{D\cal A}~\sum_{n=0}^{\infty}~\frac{1}{n!}~\prod_{i=1}^{n}
\Big[\int_{\frac{1}{\Lambda^{2}}}^{\infty}{\frac{d\alpha_{i}}{\alpha_{i}}}}
\int_{x(0)=x(\alpha_{i})}~Dx~{e^{\int_{0}^{\alpha_{i}}{d\tau [N(\tau_{i})]}}}\Big]
e^{-\int{G(x)}d^{3}x}\nonumber\\
\end{eqnarray}
  Here $N(\tau_{i})$  and G(x) are $$N(\tau_{i})={\cal
H}(\tau_{i})+\frac{1}{2}\dot
x^{\mu}_{i}\partial^{\rho}B_{\rho\mu}(\tau_{i})-i\dot
x^{\mu}_{i}A_{\mu}(\tau_{i})~~\hbox{and}$$  $$G(x)=iC_{\mu\nu}(y^{\mu\nu}+B^{\mu\nu})+\chi^{2}+\frac{i}{4\pi}\epsilon_{\mu\nu\lambda}
A_{\mu}\partial_{\nu} A_{\lambda}$$ respectively and the measure  $D{\cal A}=DADBDCD\chi$.
\section{Gauge field integration}
 We first rewrite
$\int_{x}{C_{\mu\nu}y^{\mu\nu}_{\theta}}$ as $\int_{x}{\Gamma_{\theta}^{\nu}A_{\nu}}$ (after omitting surface terms) where $\Gamma_{\theta}^{\nu}$ is
\begin{equation}
\Gamma_{\theta}^{\nu}=[-\theta^{\mu\alpha}\partial_{\alpha}C_{\mu}^{~\nu}+\theta^{\mu\nu}\partial^{\sigma}C_{\mu\sigma}-\frac{1}{2}\theta^{\alpha\nu}\partial_{\alpha}C^{\gamma}_{~\gamma}].
\end{equation}
For gauge field integration we collect all the $A_{\mu}$ terms in the above partition function and write them as

$$e^{-i\int{d^{3}x[\frac{1}{4\pi}A^{\mu}d_{\mu\nu}A^{\nu}]-A_{\mu}(x)(-J^{\mu}+\Gamma_{\theta}^{\mu})]}}$$
where we have used the  definition for the particle current, the
current associated with the particle moving along the Wilson loop,
as
\begin{equation}
J_{\mu}=\int{d\tau\dot x_{\mu} {d\tau}\delta^{3}(x-x^{c}(\tau))}
\label{pcu}
\end{equation}
and
$d_{\mu\lambda}=\epsilon_{\mu\nu\lambda}\partial^{\nu}$.
Note that unlike in the commutative case, in the absence of Chern-Simon
term, particle current is non-vanishing. After the gauge field
integration (omitting $\theta^{2}$ terms) the partition function
become
\begin{eqnarray}
Z&=&\int{D\Omega~\sum_{n=0}^{\infty}~\frac{1}{n!}~\prod_{i=1}^{n}\Big[\int_{\frac{1}{\Lambda^{2}}}^{\infty}{\frac{d\alpha_{i}}{\alpha_{i}}}}\int_{x(0)=x(\alpha_{i})}D
x e^{-S_{1}}{e^{-\int_{0}^{\alpha_{i}}{d\tau [\omega_{i}]}\Big]}}.
\label{afA}
\end{eqnarray}
In the above $$\omega_{i} = {\cal
H}_{i}-2\pi\Gamma_{\theta}^{\mu}(d^{-1})_{\mu\rho}J^{\rho}_{i}+L_{1}^{i}~~\hbox{and~the~measure~}D\Omega=DBDCD\chi. $$
 Where we have used
$$L_{1}^{i}={i\pi}(J^{\mu}_{i}(d_{\mu\nu})^{-1})J^{\nu}_{i}~\hbox{and}~~S_{1}=\int{d^{3}x
[iC_{\mu\nu}B^{\mu\nu}+\chi^{2}}]$$ In the above partition function
the integral $e^{-\int{d\tau L_{1}^{i}}}$ is of the form
\begin{equation}
e^{-i\pi\int{d^{3}xJ_{\mu}(d_{\mu\nu})^{-1}J_{\nu}}}=e^{{i\pi}(W(C_{n})+\sum_{i\neq
j}2n_{ij})}
\end{equation}
where $W(C_{n})$ is the writhe of the curve $C_{n}(=\bigcup
C_{i})$ and $n_{ij}$ is the linking number of the curves $C_{i}$
and $C_{j}$\cite{poly}. The linking number term does not contribute. The
writhe $W(C_{n})$ has the expression
$W(C_{n})=\Omega(C_{n})+2k+1$, where $\Omega(C_{n})$is the
Polyakov factor\cite{poly}, and 2k+1 is an odd integer. Thus
\begin{equation}
e^{-\pi iW(C_{n})}=(-1)e^{-i\frac{1}{2}\Omega(C_{n})}.
\label{phase}
\end{equation}
 The coefficient of $W(C_{n})$ is dictated by the choice of the coefficient of Chern-Simon term.
 Interestingly this encodes both spin and statistics of the transformed field. The (-1) in the above equation is responsible for the expression appearing as determinent rather its inverse (see Eqn.(\ref{det}) below ), which leads to Grassmanian nature of the transmuted field. The coefficient $\frac{1}{2}$ of Polyakov's spin factor is responsible for the spin $\frac{1}{2}$ nature of the transmuted field through the well known properties of spin $\frac{1}{2}$ coherent states \cite{peril}. Using Eqn.(\ref{phase}), the partition function becomes.
\begin{equation}
Z=\int{D\Omega
e^{-\int{d^{3}x(iC_{\mu\nu}B^{\mu\nu}+\chi^{2})}}}
e^{-\int{\frac{d\alpha}{\alpha}\int{ Dx(\tau)}e^{-\int_{0}^{\alpha}{d\tau[\tilde M]+(-1)i\frac{1}{2}\Omega-iV_{\mu}J^{\mu}}}}} \label{su2}
\end{equation}
where $$\tilde M= \frac{1}{4}(\delta^{\mu\nu}-B^{\mu\nu})\dot
x_{\mu}\dot x_{\nu}+V(\tau_{i})$$ and $V_{\mu}$ is given by
\begin{eqnarray}
V_{\mu}&=&[2\Gamma_{\theta}^{\sigma}(d_{\sigma\mu})^{-1}+\frac{i}{2}\partial^{\rho}B_{\rho\mu}]
\label{compv}
\end{eqnarray}
The addition of  Polyakov spin factor to the path integral for
spinless particle both in free and  in the presence of background
scalar and vector fields have been studied in \cite{sivtg}.
Following this procedure, we obtain
\begin{eqnarray}
&&-\int_{\Lambda^{-2}}^{\infty}\frac{d\alpha}{\alpha}\int { Dx(\tau)} e^{-\int_{0}^{\alpha}{d\tau[\frac{1}{4}(\delta^{\mu\nu}-B^{\mu\nu})\dot
x_{\mu}\dot x_{\nu}+V)]+(-1)i\frac{1}{2}\Omega-i\oint
V_{\mu}dx^{\mu}}}\nonumber\\&&=(-1)\int_{\Lambda^{-2}}^{\infty}\frac{d\alpha}{\alpha}Tr
e^{-\alpha[\frac{{\cal D}}{{\cal A}}+\tilde V+M_{F}]} \label{sir}
\end{eqnarray}
where $\Lambda$ is cut-off. This makes use of the well known result
\begin{equation}
\int_{{\hat n}(0)={\hat n}(\l)}{\cal D}{\hat n} ~e^{i
\int_{0}^{\l} d{\tau}(H({\hat n})+ \frac{1}{2}\Omega(\hat
n))}=Tr\left<{\hat n}| ~e^{{i} H(\tau_\mu)}|{\hat n}\right>
\end{equation}
where ${\hat n}$ are the $SU(2)$ coherent states and $\tau_\mu$
are the Pauli matrices. Here we define
\begin{eqnarray}
&&{\cal D}=(i\partial_{\mu}-V_{\mu})\tau^{\mu},~{\cal
A}=\sqrt{det(\delta_{\mu\nu}-B_{\mu\nu})},\\&& \tilde
V=-\frac{\sqrt\pi}{4\Lambda}\Big[(m^{2}B^{\mu}_{~\mu}+2\sqrt{\lambda}\chi(x)(1-B^{\mu}_{~\mu}))]\\&&
{\hbox{and}}~M_{F}=\frac{\sqrt\pi}{4\Lambda}(m^{2}+\Lambda^{2} ~ln
2)
\end{eqnarray}
where $M_{F}$ is mass of the Fermion.
  Using the above result in (\ref{su2}) we get
\begin{equation}
Z=\int{DBDCD\chi~{e^{-\int{d^{3}x(C_{\mu\nu}B^{\mu\nu}+\chi^{2})}}}}
~det~\Big[\frac{{\cal D}}{{\cal A}}+ \tilde V+M_{F}\Big]
\label{det}
\end{equation}
Note that -1 in Eqn.(\ref{sir})is responsible for the determenant
to appear in the numerator.
 This can be written as functional integral over fermionic fields
and then integrating over $\chi$, we get the partition function as
\begin{eqnarray}
Z=\int{DBDCD\Psi
D\bar\Psi~{e^{-\int{d^{3}x(C_{\mu\nu}B^{\mu\nu})}}}}e^{-\int{d^{3}x\bar\Psi\big[\frac{2{\cal
D}}{{\cal A}}+\tilde V_{1}+M_{F}\big]\Psi}-g(\Psi\bar\Psi)^{2}}
\label{final}
\end{eqnarray}
 where
$$\tilde V_{1}=-\frac{\sqrt\pi}{4\Lambda}\big[m^{2}B^{\mu}_{~\mu}]~~\hbox{and}~~
g(x)=\frac{\pi\lambda}{16\Lambda^{2}}(1-B^{\mu}_{~\mu}(x))^{2}.$$
This result is non perturbative in $\lambda$ and the interacting
fermionic theory is non-local. When
$\lambda\rightarrow0$ the theory continues to be non-local.
 This theory is different from the theory derived from
a naive generalization of Fermionic theory one obtains in NC space-time by expanding $\ast$
product to first order in $\theta$. Such theory will not have non-locality. In the limit
$\theta\rightarrow0$, Lagrangian in Eqn.(\ref{final}) becomes
\begin{eqnarray}
L&=&\int{d^{3}}x ~(C_{\mu\nu}B^{\mu\nu})+2\bar\Psi\frac{1}{{\cal A}}[i\partial_{\mu}+\frac{i}{2}\partial^{\rho}B_{\rho\mu}]\tau^{\mu}\Psi\nonumber\\
&-&\bar\Psi\frac{\sqrt\pi}{4\Lambda}[m^{2}B^{\mu}_{~\mu}]+M_{F}\bar\Psi\Psi-\frac{\pi\lambda}{16\Lambda^{2}}(1-B^{\mu}_{~\mu}(x))^{2}(\bar\Psi\Psi)^{2}.
\end{eqnarray}
  Now integration over the field $C_{\mu\nu}$ (In the partition function) set $B_{\mu\nu}$ to vanish. Hence in the case of $\theta\rightarrow 0$ but $\lambda\neq0$
the commutative result in \cite{pssmpl}, which is a local $(\Psi\bar\Psi)^{2}$
fermionic theory is retrived. When $\theta\rightarrow 0$ and
$\lambda=0$ we get the commutative result, i.e free fermion. Thus
the commutative limit is smooth.

\section{Conclusion}
In this paper we have studied Fermionization in 3 dimensional
NC space, where NC $\lambda\phi^{4}$ theory coupled to
$U_{\ast}(1)$ gauge field governed by Chern-Simon action. The dual
Fermionic partition function derived, (non-perturbative in
$\lambda$) is non-local for both $\lambda=0$ and $\lambda\ne0$ cases. As it is clear Fermionic mass term does not get
$\theta$ correction. In $\theta\rightarrow0$ limit commutative result $\cite{sssmpl,pssmpl}$ is recoverd. Note also when $m\rightarrow0$, Dirac paricle has a non zero mass (dependent on cut off $\Lambda$). This is expected as Chern-Simon term in the Bosonic theory
is parity violating, which is reflected in the non zero mass of the fermionic theory. In the NC case it is possible for a real scalar to couple with gauge field (unlike in the commutative case). Thus it is natural to seek Fermionisation of real scalar coupled to Chern-Simon term. The SW mapped action for real scalars is (up to ordrt $\theta$)
\begin{eqnarray}
\label{action_scalar_field0} S_{\varphi} = \frac{1}{2} \int d^4x \,
\left[ \partial^\mu \varphi
\partial_\mu \varphi + 2 \theta^{\mu\alpha} {F_\alpha}^\nu \left(
-
\partial_\mu \varphi
\partial_\nu \varphi + \frac{1}{4} \eta_{\mu\nu} \partial^\rho \varphi
\partial_\rho \varphi \right) \right].
\end{eqnarray}
Here the coupling to gauge field is through non-minimal coupling only. For the application of Poliyakov's approach it is necessary to have Wilson loop term (i.e, minimal coupling), which is absent here. Hence, straight forward extension of 
this procedure to real scalar is not possible. It is an interesing problem to see how the real scalar can be fermionised.\\\\
{\bf Acknowledgement}
KMA and MS acknowledges DST for support through a project.

\end{document}